\documentclass[journal=jpccck,manuscript=article]{achemso}

\usepackage[version=3]{mhchem} % Formula subscripts using \ce{}

\usepackage{xcolor}
\usepackage{amsmath}
\usepackage{amssymb}
\author{Matheus Jacobs}
\affiliation
{Physics Department and IRIS Adlershof, Humboldt-Universit\"at zu Berlin, 12489 Berlin, Germany}
\email{jacobs@physik.hu-berlin.de}

\author{Jannis Krumland}
\affiliation
{Physics Department and IRIS Adlershof, Humboldt-Universit\"at zu Berlin, 12489 Berlin, Germany}

\author{Ana M. Valencia}
\affiliation
{Physics Department and IRIS Adlershof, Humboldt-Universit\"at zu Berlin, 12489 Berlin, Germany}
\alsoaffiliation{Institute of Physics, Carl von Ossietzky Universit\"at Oldenburg, 26129 Oldenburg, Germany}

\author{Caterina Cocchi}
\affiliation
{Physics Department and IRIS Adlershof, Humboldt-Universit\"at zu Berlin, 12489 Berlin, Germany}
\alsoaffiliation{Institute of Physics, Carl von Ossietzky Universit\"at Oldenburg, 26129 Oldenburg, Germany}
\altaffiliation{Center for Nanoscale Dynamics  (CeNaD), Carl von Ossietzky Universität, 26129 Oldenburg, Germany}

\email{caterina.cocchi@uni-oldenburg.de}

\title{Pulse-Induced Dynamics of a Charge-Transfer Complex from First Principles}
\begin{document}

%\begin{tocentry}

%\section*{TOC Graphic}
%\begin{figure}[H]
%    \centering
%    \includegraphics[width=8.25 cm]{toc.png}
%\end{figure}

%\end{tocentry}

\newpage

\begin{abstract}
The ultrafast dynamics of charge carriers in organic donor-acceptor interfaces are of primary importance to understanding the fundamental properties of these systems. In this work, we focus on a charge-transfer complex formed by quaterthiophene $p$-doped by tetrafluoro-tetracyanoquinodimethane and investigate electron dynamics and vibronic interactions also at finite temperatures by applying a femtosecond pulse in resonance with the two lowest-energy excitations of the system with perpendicular and parallel polarization with respect to the interface.
The adopted \textit{ab initio} formalism based on real-time time-dependent density-functional theory coupled to Ehrenfest dynamics enables monitoring the dynamical charge transfer across the interface and assessing the role played by the nuclear motion. 
Our results show that the strong intermolecular interactions binding the complex already in the ground state influence the dynamics, too.
The analysis of the nuclear motion involved in these processes reveals the participation of different vibrational modes depending on the electronic states stimulated by the resonant pulse.
Coupled donor-acceptor modes mostly influence the excited state polarized across the interface, while intramolecular vibrations in the donor molecule dominate the excitation in the orthogonal direction.
The results obtained at finite temperatures are overall consistent with this picture, although thermal disorder contributes to slightly decreasing interfacial charge transfer.
\end{abstract}

\newpage
\section{Introduction}
The photophysics of organic donor-acceptor complexes is an attractive field of research to unfold the potential of these systems as active components for optoelectronic devices~\cite{schw+14afm,hool+15jpcb,fuze+16jpcl,neel+18jpcl,xu+18oe,cui+21jpcl,theu+21jpcc}.
The complex landscape of often intertwined electronic, optical, and vibrational excitations characterizing this material class depends crucially on the structure-property relationships of the specific systems~\cite{salz+12prl,gao+13jmcc,dinu+15natcom,neel+18jpcl,beye+19cm} and on the characteristics of the building blocks~\cite{salz+16acr,arvi+20jpcb,mans+20jmcc}. 
Different doping mechanisms emerge according to the electronic interactions between donor and acceptor molecules and crucially influence the response of the complex to external stimuli. 

The creation of an anion and a cation upon photoexcitation, known as ion pair formation~\cite{salz+16acr}, is an effective way to transfer charge carriers across the interface~\cite{mend+15ncom,salz+16acr}.
Examples of such systems are $p$-doped thiophene polymers~\cite{mend+15ncom,burk+19am} which are often used as active materials in organic solar cells~\cite{berg-kim18jrse}.
Another class of donor-acceptor compounds is given by so-called charge-transfer complexes (CTC)~\cite{mend+15ncom,salz+16acr}.
These doped systems are characterized by fractional charge transfer in the ground state accompanied by electronic hybridization between the frontier orbitals of the two molecular species~\cite{aziz+07am,salz+12prl,mend+13acie,mend+15ncom,vale-cocc19jpcc}. 
In the excited state, different types of excitations can be formed, again, depending on the characteristics of the building blocks~\cite{vale-cocc19jpcc} and on the local interfaces between them~\cite{valencia+PCCP+20} even in the crystalline phase~\cite{guer+21jpcc}.
Notably, in CTC, the lowest-energy excitation is bright owing to the non-negligible overlap between the bonding and anti-bonding frontier orbitals~\cite{mend+15ncom,vale-cocc19jpcc,valencia+PCCP+20}.

Understanding the excitation dynamics of donor-acceptor complexes is of prime importance to clarify the fundamental processes ruling photoabsorption, charge separation, and diffusion~\cite{rozz+13natcom,falk+14sci,dorf+22jpm}: this is essential information to predict the performance of these systems in optoelectronic devices and solar cells~\cite{tama+17nano,chen+20jpcl,bolz+21jpcl}.
Exploring the ultrafast regime of excitation in the natural, sub-picosecond timescale of electrons gives insight into the charge-transfer mechanisms driven by resonant photon absorption~\cite{desi-lien17pccp}.
The coherent coupling between electronic and vibrational degrees of freedom was identified as the key mechanism driving the separation of photoexcited charge carriers across the interface of a non-covalently bound donor-acceptor blend~\cite{falk+14sci,tamu+08jpcb,polk+17jpb,bind+18prl,popp+19jpcl,peng+22jpcl}.
Moreover, time-resolved Raman spectroscopy studies revealed the relevance of vibronic motion in driving chemical reactions and electron transfer in donor-acceptor compounds~\cite{hoff+14jpca,elli+18jpca}.
Stimulated by these findings, it is interesting to turn to CTC and try to answer fundamental questions for such systems characterized by frontier-orbital hybridization and fractional charge transfer in the ground state.
Addressing (i) the characteristics of ultrafast charge transfer upon the resonant excitation of different electronic transitions, (ii) the role of vibronic coupling -- particularly, which vibrational modes come into play under specific excitation conditions -- and (iii) the effects of temperature on the coherence of the process are of primary importance to gain deeper understanding on the intrinsic light-matter coupling processes occurring in this class of organic materials.

In this work, we focus on the prototypical CTC formed by a quaterthiophene (4T) molecule $p$-doped by the strong electron acceptor 2,3,5,6-tetrafluoro-7,7,8,8-tetracyanoquinodimethane (F4TCNQ).
We investigate the charge-carrier dynamics in this system, including the nuclear motion, adopting state-of-the-art first-principles methods. 
In the framework of real-time time-dependent density-functional theory coupled with the Ehrenfest scheme, we monitor the distribution of the electron density across the resonantly excited interface in a 100-fs time window.
We analyze the vibrational modes that participate in the dynamics and explore temperature effects adopting thermalized ensembles from Born-Oppenheimer molecular dynamics simulations. 
Our results reveal substantially different scenarios depending on the polarization of the resonant radiation in both the electronic and the vibronic dynamics.
The excitation of the transition between the bonding and anti-bonding frontier orbitals leads to pronounced oscillations of the charge distribution.
The charge density sloshes between the donor and the acceptor molecule with the participation of coupled vibrational modes of both 4T and F4TCNQ.
In contrast, the resonant stimulation of the second excited state of the system, corresponding to an optical transition polarized perpendicular to the interfacial direction, is dominated by the dynamics in the 4T.
In this scenario, the charge distribution does not slosh between donor and acceptor but under the simulated conditions, charge transfer does not effectively increase with respect to the ground state either.
At finite temperatures, disorder effects become prominent: the linear absorption spectrum becomes broader and the external pulse leads to the (partly off-resonant) activation of the first and second excitations simultaneously. 
The results obtained in the adopted formalism, where quantum effects are accounted for only for the electronic system, point to an overall decrease in charge transfer compared to the ground state.

%%%%%%%%%%%%%%%%%%%%%%%%%%%%%%%%%
\section{Methods}

\subsection{Theoretical Background}

The results presented in this work are obtained from first principles in the framework of real-time time-dependent density-functional theory (RT-TDDFT). 
In this approach, the time-dependent Kohn-Sham (TDKS) equations~\cite{rung-gros84prl},
\begin{equation}
i\frac{\partial}{\partial t}\phi_i(\textbf{r},t) =\left( -\frac{\nabla_{\mathbf{r}}^2}{2}+v_{\text{KS}}[n](\textbf{r},t)\right)\phi_i(\textbf{r},t),
\label{eq:TD-KS}
\end{equation}
are solved directly through their propagation in real time.
In Eq.~\eqref{eq:TD-KS}, $\phi_{i}(\textbf{r},t)$ are the wave functions that are used to compute the time-dependent density, $n(\mathbf{r},t)=\sum_i P_i|\phi_i(\mathbf{r},t)|^2$, where $P_{i}$ is the population of the $i$-th state.  The central quantity in Eq.~\eqref{eq:TD-KS} is the TDKS potential defined as
\begin{equation}
\begin{aligned} 
v_{\text{KS}}[n](\textbf{r},t) = v_{\text{ions}}(\textbf{r},t) + v_{\text{ext}}(\textbf{r},t) + \int\text d^3r'\frac{n(\textbf{r}',t)}{|\textbf{r}-\textbf{r}'|} + v_{\text{xc}}[n](\textbf{r},t),
\end{aligned}
\label{eq:TDDFT potential}
\end{equation}
where $v_{\text{ions}}(\textbf{r},t)$ describes the interactions between electrons and nuclei, $v_{\text{ext}}(\textbf{r},t)$ is the external time-dependent potential, the integral is the Hartree potential, and $v_{\text{xc}}[n](\textbf{r},t)$ corresponds to the exchange-correlation (XC) potential. 
To describe electron-nuclear couplings, RT-TDDFT is interfaced with the Ehrenfest scheme, a non-adiabatic approach for molecular dynamics~\cite{marq+03cpc,rozz+17jpcm} where the nuclei are treated classically and average electrostatic forces are calculated for the electrons as
\begin{equation}
\begin{aligned} 
M_{i}\ddot{ \textbf{R}}_{i}=-\sum_{j=1}^{N} \langle \phi_{j}\left|\frac{\partial v_{KS}}{\partial \textbf{R}_{i}}\right|\phi_{j} \rangle,
\end{aligned}
\label{eq:E-TDDFT}
\end{equation}
where $\phi_j$ are the solutions of Eq.~\eqref{eq:TD-KS}, $M_{i}$ the mass of the $i$-th nucleus, and $\textbf{R}_{i}(t)$ its trajectory.
The solution of Eq.~\eqref{eq:E-TDDFT} allows monitoring the kinetic energy of each normal mode of the system through the relation
 \begin{equation}
    E^{\text{kin}}_\alpha(t) = \dfrac{\dot{\cal Q}_\alpha^2(t)}{2},
 \end{equation}
where $\dot{\cal Q}_\alpha$ is the time derivative of the normal coordinate associated with the vibrational mode $\alpha$. 

To explore the dynamics in different thermal configurations, we performed Born-Oppenheimer molecular dynamics (BOMD) simulations~\cite{marx_hutter_2009}. In this scenario, the system is coupled to a thermal bath described by the Nose thermostat~\cite{nose+JPC+84,hoover+PRA+85} until thermalization is reached for each target temperature (100 K, 200 K, and 300 K). We choose such a thermostat to ensure proper sampling of the canonical ensemble with thermal fluctuations.\cite{huenenberger+philippeAdvComp2005,brau+18jctc}
No quantum effects are taken into account in the molecular dynamics calculations. While this is a common strategy also consistent with the classical approximation taken for the nuclei, we give to consider that this can be expected to somewhat underestimate the effect of nuclear motion on the electronic dynamics due to the lack of zero-point energy~\cite{barbattiKakali2016}. However, we do not expect this to influence our results on a qualitative level. 
Another caveat concerns the inability of Ehrenfest dynamics to describe wave packet splitting.
As a consequence, results can become inaccurate if excited-state dynamics are governed by potential-energy surfaces that differ qualitatively from the ground-state one, \textit{e.g.}, that do not support the bound nuclear motion and would lead to dissociation. Here, we checked that no substantial reconfiguration occurred between the molecules, which remain in a face-to-face mutual arrangement due to their strong coupling in the CTC.

To sample uncorrelated snapshots, we compute the autocorrelation function of the energy as 
\begin{align}\label{eq.ACF}
    {\rho(E_{i},E_{i+k})} = \frac{E_{i} \ast E_{i+k}}{s(E)^{2}},
\end{align}
where $E_{i} \ast E_{i+k}$ represents the convolution between energies $E_{i}$ and $E_{i+k}$, $s(E)$ is the energy variance, and $k$ is the time step between two consecutive samples. The snapshots are sampled at intervals based on $\lim{\rho(E_{i},E_{i+k}) \to \ 0}$, which can be interpreted as the time required by the system to lose the memory of its previous configuration. Corresponding positions and velocities are used to initialize the RT-TDDFT simulations at each temperature.
The time evolution of different quantities $X$ ensuing laser excitation is averaged across the ensemble. The result is closely linked to the time-dependent dipole-$X$ correlation function, since the laser is coupled to the system in the dipole approximation.

Linear-response TDDFT calculations~\cite{casi95,casi96tcc} are performed to analyze the character of the optical excitations, including their polarization, oscillator strengths, and composition.
In this framework, the transition density is computed as 
\begin{equation}
    \rho_{TD} (\textbf{r})=\sum_{i,j} A_{ij}^{\lambda} \phi_{i}^*(\textbf{r}) \phi_{j}(\textbf{r}),
    \label{eq:rho_TD}
\end{equation}
where $A_{ij}^{\lambda}$ are the coefficients associated with each single-particle transition contributing to the excitation $\lambda$.
Eq.~\eqref{eq:rho_TD} provides a visual representation of the orientation of the transition dipole moment of each excitation.
Density functional perturbation theory calculations~\cite{blum+09cpc,shan+18njp} were carried out to compute vibrational frequencies and normal modes in the ground-state electronic configuration.

%%%%%%%%%%%%%%%%%%%%%%%%%%%%%%%%
\subsection{Computational Details}

RT-TDDFT+Ehrenfest calculations are performed using the OCTOPUS code~\cite{tanc+20jcp}. The FIRE algorithm~\cite{bitz+06prl} is employed to relax the initial structure until the residual forces are smaller then $10^{-3}$~eV/\AA{}. For the ground-state calculations, a real-space grid with spacing 0.11~\AA{} is adopted in a simulation box formed by interlocked spheres of radius 6.0~\AA{} centered on each atom. Hartwigsen-Goedecker-Hutter norm-conserving pseudopotentials simulate core electrons~\cite{hartwigsen1998relativistic,goedecker1996separable}. The local-density approximation (LDA) in the Perdew-Zunger parametrization~\cite{perd-wang92prb} is used for the XC potential.
We checked that this functional is able to deliver a qualitatively correct description of the linear absorption properties of the considered CTC in comparison with many-body perturbation theory calculations~\cite{vale-cocc19jpcc}.
Also, the adiabatic LDA in combination with Ehrenfest dynamics offers a reliable and affordable way to compute vibronic dynamics~\cite{rozz+13natcom,falk+14sci,krumland+2020jcp} in spite of its known limitations~\cite{ragh-nest11jctc,fuks+2011prb,fuks-mait+14pccp}.
We use the term ``vibronic'' herein implying with it the coupling between optically driven electronic excitations and vibrational motion with the latter computed at the level of classical molecular dynamics, as discussed above. This being said, we acknowledge that part of the community uses this term primarily in the context of electron-vibration coupling problems where nuclei are treated at the quantum level~\cite{polk+18ijqc}.

To compute the linear absorption spectrum according to the Yabana-Bertsch scheme~\cite{yabana1996time}, the KS orbitals obtained by solving Eq.~\eqref{eq:TD-KS} are propagated through the approximated enforced time-reversal symmetry (AETRS) propagator~\cite{castro2004propagators} with a time step of 0.37~as and a total simulation duration of 20~fs. A ``kick'' of magnitude 0.0053~\AA{}$^{-1}$ is applied in all Cartesian directions to trigger the time propagation. The grid parameters are the same as in the ground-state calculations.
Absorption spectra computed at finite temperatures are averaged over 100 configurations.
In the ultrafast dynamics simulations at 0~K, a time-dependent electric field is applied in resonance with two selected excitations giving rise to absorption maxima at 1.2~eV and 1.5~eV in the linear absorption spectrum. The applied laser pulse of Gaussian shape is centered at 8~fs and has a standard deviation of 2~fs. 
The peak intensity is set to 300~GW/cm$^{2}$ in order to simulate the response of the system to a strong laser, thus mimicking conditions similar to those achieved in specialized labs~\cite{borz+10oe,pico+11pra}.
Such intensities are typically adopted in RT-TDDFT simulations of organic, inorganic, and hybrid materials~\cite{degi+13chpch,krumland+2020jcp,yama+18prb,zhan+20pla,jaco+20apx}
In previous work on an organic/inorganic interface~\cite{jaco+22acsanm}, we found that the same intensity ensures effective excitation without inducing sizeable nonlinear effects or the ionization of the system.
On the other hand, we checked that adopting a much weaker pulse may give rise to spurious numerical noise, which, in turn, may affect the quality of the results and the reliability of their interpretation.
We emphasize that by exciting the system with a resonant pulse, we aim to investigate the effects of electronic coherence on the nuclear motion of the CTC. 
Such a scenario cannot be realized if the system is initially prepared in an excited state~\cite{negr+12jpcc,ovie+12jpcl,domi-frau21pccp}. 
We emphasize that our simulations do not target sunlight harvesting or photosynthesis, which are driven by incoherent and broadband radiation of much lower intensity compared to the one applied here.

BOMD simulations are performed with the CP2K package~\cite{kueh+20jcp}. The computational parameters adopted in this step are the same as for the RT-TDDFT calculations, with the exception of the adopted basis set (DZVP~\cite{VandeVondele+juerg+jcp+07}) including a plane-wave cutoff of 600~Ry.
In this analysis, we considered finite temperatures of 100~K, 200~K, and 300~K for which thermal effects are expected to be sizeable; lower temperatures, for which quantum nuclear effects are known to be even more prominent than at room temperature~\cite{ross21jcp,krum+22prb}, 
are deliberately excluded from this investigation.
During thermal equilibration at 100~K, 200~K, and 300~K, a heating ramp drives the system smoothly to the target temperature in the canonical ensemble. To sample structures without any external constraints, at each temperature the system is propagated for 3~ps in the microcanonical ensemble (NVE). 
Each snapshot obtained from this pre-processing is then propagated for 100~fs in the framework of RT-TDDFT+Ehrenfest (AETRS propagator) adopting the OCTOPUS code and the same parameters listed above. 
To evaluate charge transfer, the density is output every 2~fs and the Bader charges are calculated~\cite{bade90} averaging over 100 different structures in each thermal ensemble.

%%%%%%%%%%%%%%%%%%%%%%%%%%%%%%%%%%%%%%%%%%%%%%%%%%%
\section{Results and Discussion}

\subsection{Ground-State Charge Transfer and Linear Absorption}

4T-F4TCNQ is a CTC characterized by fractional charge transfer in the ground state~\cite{vale-cocc19jpcc,mend+15ncom}.
This behavior is induced by the strong $p$-doping character of F4TCNQ which according to our results (see Table~\ref{tab:table_ct}) withdraws more than 0.6~$e$ from the donor.
This estimate is in line with reference values in the literature obtained at analogous levels of theory~\cite{valencia+PCCP+20}; quantitative differences are ascribed to the adopted approximations (basis sets, XC functional, etc.).
Structural details in the construction of the complex and in particular local interfaces between donor and acceptor molecules may play a role, too, as extensively discussed in Ref.~\citenum{valencia+PCCP+20}.
However, it is generally accepted that the bimolecular cluster considered in this analysis is a reasonable model for 4T-F4TCNQ pseudocrystals and blends~\cite{zhu+11cm,mend+15ncom}.
We acknowledge, though, that long-range effects may not be captured properly in such a configuration~\cite{guer+21jpcc} and the extent of charge delocalization may be underestimated with respect to experiments. However, measurements for the specific system and conditions under consideration are currently unavailable and, thus, a direct assessment is not possible to date.
Examining the values reported in Table~\ref{tab:table_ct} for finite temperatures, we notice that fluctuations with respect to the result at 0~K are below 10$\%$.
The largest variation, on the order of 0.05~$e$, is seen at 100 K.
At higher temperatures, the amount of charge transferred decreases again reaching very similar values (0.654 and 0.651~$e$) at 200~K and 300~K, respectively.
This behavior suggests that changes with respect to the 0~K scenario are induced by the thermal fluctuations that are naturally present in the simulations. However, the overall charge transfer, which is driven by the local interactions between the donor and the acceptor species~\cite{valencia+PCCP+20}, is robust enough to be almost independent of temperature.

\begin{table}[h]
\centering
\begin{tabular}{cc}
\hline \hline
\multicolumn{1}{c}{Temperature (K)} & Charge transfer (e) \\ \hline \hline
\multicolumn{1}{c}{0}               & 0.622      \\ \hline
\multicolumn{1}{c}{100}             & 0.674  $\pm 0.002$    \\ \hline
\multicolumn{1}{c}{200}             & 0.654  $\pm 0.001$    \\ \hline
\multicolumn{1}{c}{300}             & 0.651  $\pm 0.003$    \\ \hline \hline
\end{tabular}
\caption{\label{tab:table_ct} Absolute value of charge transfer and standard deviation in the 4T-F4TCNQ complex computed at each temperature using the Bader scheme averaging over 100 different structures.}
\end{table}

\begin{figure*}
\includegraphics[width=\textwidth]{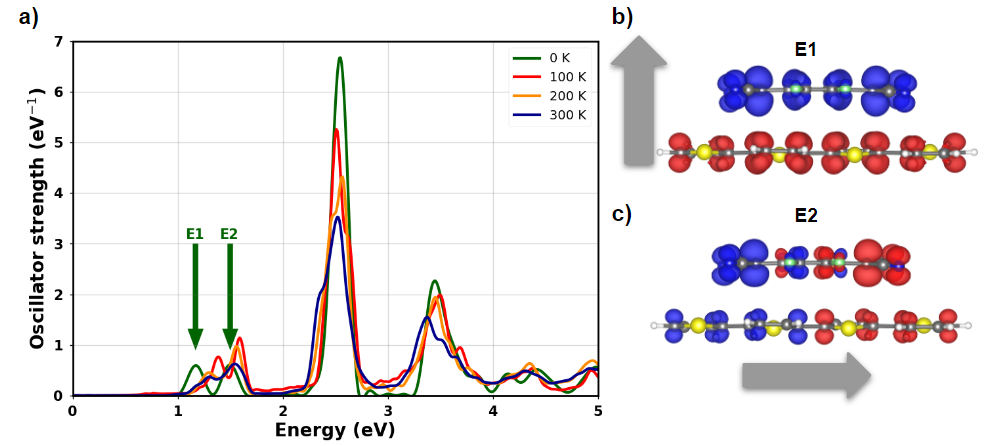}
\caption{a) Linear absorption of 4T-F4TCNQ computed at increasing temperatures. 
Transition densities associated with b) the first excitation (E1) and c) the second excitation (E2) indicated in the spectrum at 0K in panel a). Blue and red isosurfaces indicate domains of charge depletion and accumulation, respectively. The gray arrows highlight the direction of the transition dipole moment.}
\label{fig:t_density}
\end{figure*}

In the next step of our analysis, we inspect the linear absorption spectrum of the 4T-F4TCNQ complex.
The result obtained at 0~K (Figure~\ref{fig:t_density}a) is computed from RT-TDDFT in the adiabatic local density approximation.
For this reason, excitation energies and relative oscillator strengths differ quantitatively with respect to the many-body perturbation theory results of Refs.~\citenum{vale-cocc19jpcc, valencia+PCCP+20}. 
Yet, the main characteristics of those spectra are retained: two weak excitations appear in the low-energy region followed by more intense resonances at higher energies. 
The first two maxima, labeled E1 and E2 in Figure~\ref{fig:t_density}a, correspond to
Frenkel excitons in the whole complex~\cite{vale-cocc19jpcc,valencia+PCCP+20}. This nomenclature is adopted based on the evidence that the corresponding electron and hole densities are delocalized on both donor and acceptor molecules, as extensively discussed in Refs.~\citenum{vale-cocc19jpcc, valencia+PCCP+20}.
The spatial distribution of the electron-hole pairs is directly related to the orbitals contributing to them~\cite{vale-cocc19jpcc}.
%with both the electron and the hole delocalized on both molecules.
E1 stems from the transition between the highest-occupied molecular orbital (HOMO) and the lowest-unoccupied one (LUMO), bearing bonding and anti-bonding character (see Figure~S2), respectively~\cite{vale-cocc19jpcc,mend+15ncom}; as such, E1 is polarized across the donor/acceptor interface as illustrated by the transition density shown in Figure~\ref{fig:t_density}b.
We stress that the color domains in those plots refer to positive and negative values of the transition dipole moment, not to the distribution of electrons and holes.
Conversely, E2 corresponds to a transition between the HOMO-1 and the LUMO~\cite{vale-cocc19jpcc}, with the resulting dipole moment parallel to the long molecular axis (see Figure~\ref{fig:t_density}c). 
The intense maximum centered at about 2.5~eV (see Figure~\ref{fig:t_density}a) corresponds to an excitation where both the hole and the electron are localized on the donor.
Given their complementary characteristics, E1 and E2 are most suited to be targeted in the analysis of resonantly driven charge-transfer dynamics.

Before moving on in this direction, it is instructive to clarify the influence of temperature on the linear absorption spectrum of 4T-F4TCNQ.
The results reported in Figure~\ref{fig:t_density}a indicate that the primary effect is a smearing of the absorption peaks, that become less intense and undergo a redistribution of their spectral weight over a wide energy range.
This broadening is related to the average performed over of 100 different configurations in the thermalized ensembles. 
Additionally, at finite temperatures, E1 is blue-shifted by a few hundred meV compared to its counterpart at 0~K.
This behavior can be explained in terms of energetic decoupling of the HOMO and the LUMO of the complex: when temperature increases, the donor and acceptor molecules increase their mutual distance by about 0.2~\AA{} compared to the 0~K scenario (see Figure~S5). As a result, the frontier levels are shifted to lower and higher energies giving rise to the increase of the optical gap testified by the blue-shift of E1 (HOMO$\rightarrow$LUMO transition).
Considering that the energy of E2 (HOMO-1$\rightarrow$LUMO) increases with temperature by only a few hundred meV (Figure~\ref{fig:t_density}a), we can infer that the HOMO is more susceptible to thermal effects than the LUMO.
In the spectra computed at finite temperatures, the blue-shift of both peaks, combined with the enlarged smearing due to ensemble averaging, gives rise to a broad absorption maximum including both E1 and E2 (see Figure~\ref{fig:t_density}a). 
At higher energies, the effect of temperature is less dramatic.
This can be explained by the fact that the intense excitation at about 2.5~eV corresponds to an intramolecular transition in the donor~\cite{krum+21pccp}: the involved orbitals are localized on 4T and, as such, are only marginally affected by the structural reorganization of the complex on account of thermal effects.

%%%%%%%%%%%%%%%%%%%
\subsection{Charge-Transfer Dynamics and Vibronic Coupling}

\begin{figure*}
\includegraphics[width=1\textwidth]{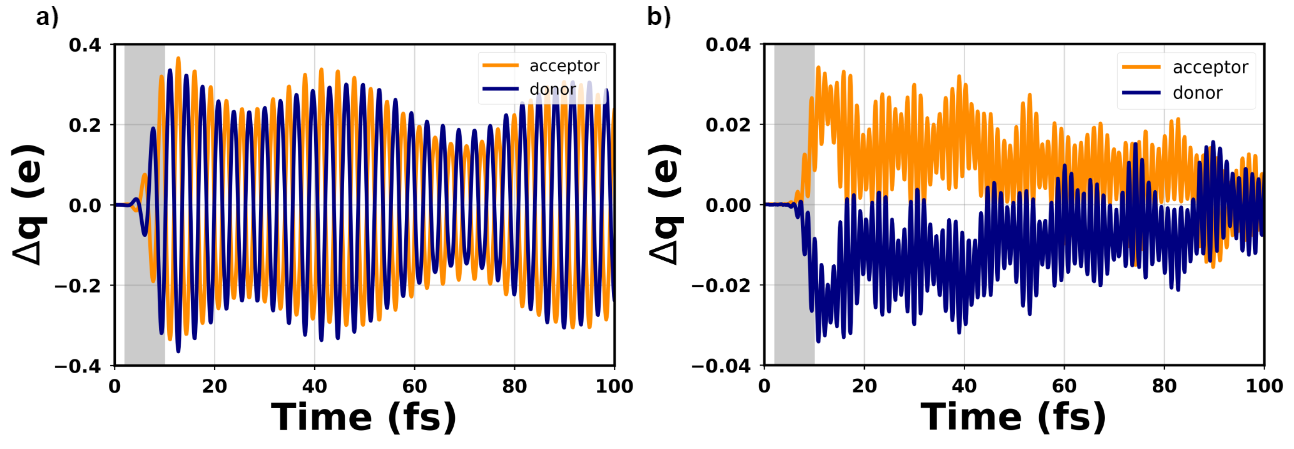}
\caption{Time evolution of the partial charges, computed with respect to the ground-state value, on the donor (4T) and acceptor molecule (F4TCNQ) of the complex upon the resonant excitation of a) E1 and b) E2. The gray, shaded area indicates the time window in which the time-dependent pulse is active.}
\label{fig:ct_dyn}
\end{figure*}

As a first step in the analysis of the charge-transfer dynamics of the CTC, we inspect the time evolution of the partial charge distribution across the interface with respect to the reference ground-state values reported in Table~\ref{tab:table_ct}.
This quantity represents the time-dependent quantum-mechanical expectation value of the charge transfer in the prepared superposition state.
By stimulating E1 at 0~K, the partial charge variation, $\Delta q$, exhibits a pronounced oscillatory behavior with a large amplitude across negative to positive values (Figure~\ref{fig:ct_dyn}a). 
The oscillation period, which remains essentially constant throughout the entire simulation window, is of approximately 3.45~fs, matching the energy of the incident pulse (1.2~eV).
This behavior indicates that the resonant pumping of E1 leads to an electronic charge sloshing across the interface that does not ultimately lead to charge separation in the considered 100-fs time window.
Vibronic motion, which is responsible for the amplitude modulation of $\Delta q$, cannot revert the mentioned trend.

The charge-transfer dynamics obtained by resonantly exciting E2 differ both qualitatively and quantitatively from the above-discussed results for E1 (Figure~\ref{fig:ct_dyn}b).
In the first 50~fs, $\Delta q$ is positive on the acceptor and negative on the donor, suggesting an enhancement of charge transfer driven by the applied resonant field. 
After about 80~fs, the partial charges on the donor (acceptor) start assuming negative (positive) values; around 100~fs, the baseline of both trends is very close to the ground-state value.
In the stimulation of E2, electron-vibrational couplings do not give rise to a clear amplitude modulation of $\Delta q$, as seen instead for E1.
Furthermore, we note that the absolute values of $\Delta q$ in Figure~\ref{fig:ct_dyn}a and Figure~\ref{fig:ct_dyn}b differ from each other by one order of magnitude.

Before continuing with the analysis, it is worth commenting on the results obtained so far in comparison with similar studies.
In Refs.~\citenum{rozz+13natcom,falk+14sci}, it was demonstrated that vibrational coherence enhances charge transfer.
However, the systems investigated in those works are intrinsically different from the one explored here.
The donor-acceptor complex addressed in Ref.~\citenum{falk+14sci} does not exhibit sizeable charge transfer nor orbital hybridization in the ground state. 
The creation of an electron-hole pair in one subsystem leads to charge separation upon the action of an external field and is assisted by vibrational motion.
Conversely, in the CTC examined here, the targeted excitations E1 and E2 are both delocalized on the entire complex (see Figure~\ref{fig:t_density}b-c and Refs.~\citenum{vale-cocc19jpcc,valencia+PCCP+20}) and are thus ``charge-transfer'' in nature.
The application of a pulse in resonance with them triggers and enhances the charge transfer with respect to the ground state with a marginal role of vibrations compared to the case discussed in Ref.~\citenum{falk+14sci}.
Also, the dynamics of E2 resemble the one obtained at the same level of theory for a prototypical hybrid inorganic/organic interface in which F4TCNQ $p$-dopes an H-terminated Si cluster~\cite{jaco+20apx}, where the pumped excitation has also perpendicular polarization with respect to the interfacial direction, \textit{i.e.}, the transition dipole moment is parallel to the long molecular axis.
Finally, the results reported in Figure~\ref{fig:ct_dyn} indicate that the laser-excited CTC remains in a coherent superposition between the ground- and the excited-state, as expected from the adopted RT-TDDFT+Ehrenfest scheme~\cite{krumland+2020jcp}.
As such, quantities of interest in the context of quantum-chemical studies on transition barriers, such as reorganization energies~\cite{zwick+08jpca,ren+13jpca,maty23pccp}, are not particularly meaningful in this context.

\begin{figure*}[h!]
\includegraphics[width=\textwidth]{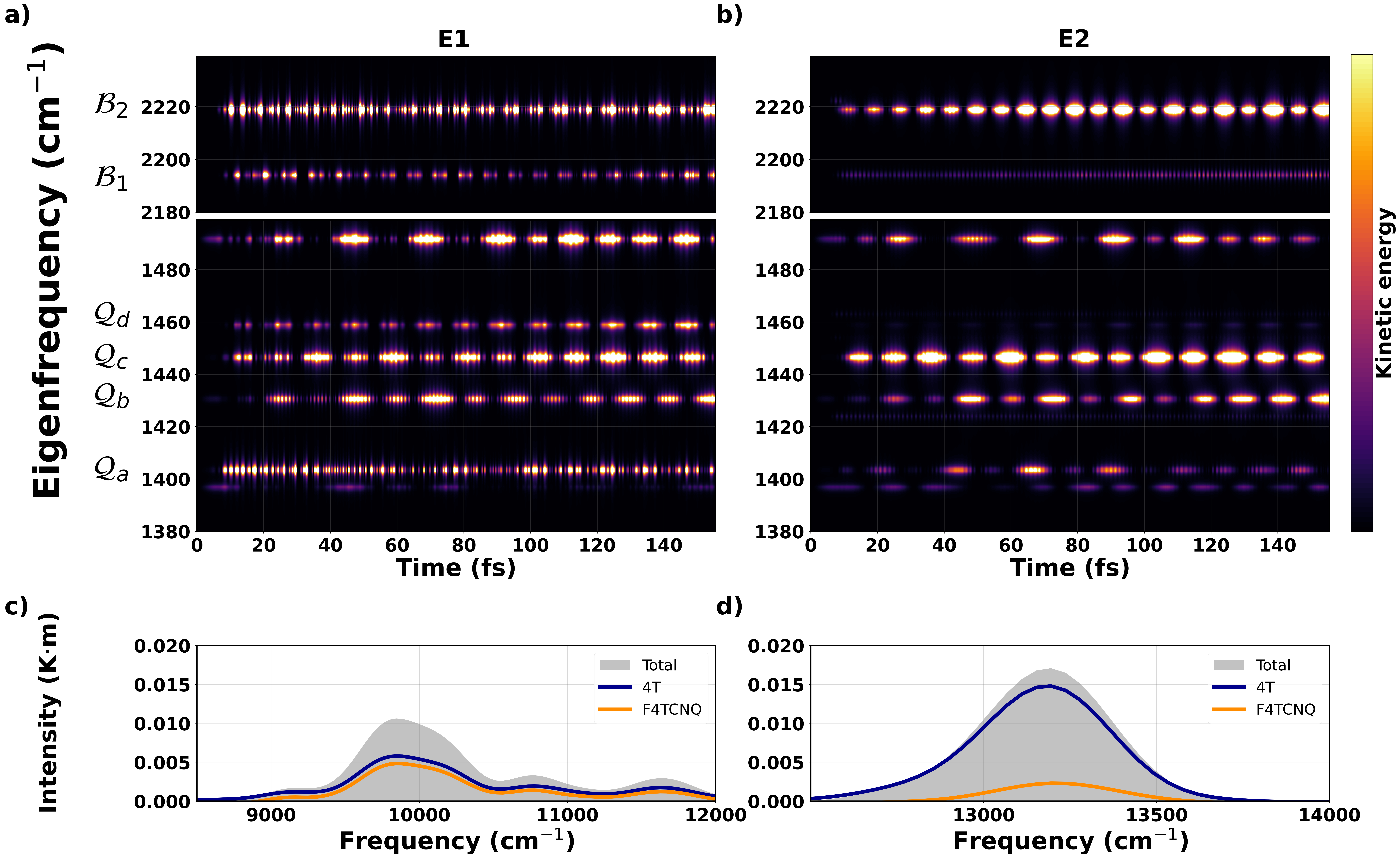}
\caption{a) Kinetic energy of the vibrational modes during the laser-induced dynamics for E1.  b) Kinetic energy of the vibrational modes during the laser-induced dynamics for E2. Power spectra of the CTC (gray area) upon the excitation of c) E1 and d) E2, with the contributions of the constituents (4T and F4TCNQ) marked by continuous lines.}
\label{fig:kin_energy}
\end{figure*}

To understand the dynamics reported in Figure~\ref{fig:ct_dyn}, it is useful to deepen the analysis on the vibronic coupling of the 4T-F4TCNQ complex.
Specifically, since both E1 and E2 are intermolecular excitations (Figure~\ref{fig:t_density}b-c), it is relevant to understand whether the vibrational modes of the acceptor and donor interact in the ground state, and if they are activated by laser excitation. 
To answer these questions, we analyze the vibrational spectrum of the complex and find three modes, with frequencies 1403 cm$^{-1}$, 1430.6 cm$^{-1}$, and 1446.5 cm$^{-1}$, coupling the C=C stretching in F4TCNQ (bonds connecting the tetrafluorinated phenyl rings with the cyano groups) with the so-called C$_\alpha$-C$_{\beta}$ mode~\cite{mans+22pccp} in 4T. 
In the following, we will refer to them as ${\cal Q}_{a}$, ${\cal Q}_{b}$ and ${\cal Q}_{c}$, respectively (see Figure~S3).
We recall that the adopted RT-TDDFT+Ehrenfest formalism includes non-adiabatic effects that are expected to be important in these dynamics. 

Following the strategy adopted in previous work~\cite{herp+21jpca}, we examine the kinetic energy accumulated by these modes during laser irradiation and the subsequent evolution of the vibronic degrees of freedom in the 4T-F4TCNQ complex.
In Figure~\ref{fig:kin_energy}a, we show a map of the kinetic energy of the normal modes during the time evolution of the system excited in resonance with E1.
All coupled modes, ${\cal Q}_{a}$, ${\cal Q}_{b}$, and ${\cal Q}_{c}$, undergo an increase in the kinetic energy; the period of their oscillation is about 3.40~fs, which corresponds to the one of the incident pulse. 
We interpret this result as follows. 
The laser polarization is perpendicular to the conjugated backbone of the molecules and hence to the covalent bonds therein. The electronic density is consequently perturbed and induces a force that drives the oscillation of the electrons in the system at the field frequency. This force interacts with the harmonic motions of some of the modes, $e.g.$, ${\cal Q}_{b}$, and ${\cal Q}_{c}$, causing their oscillation at the sum of the natural and the field frequency. We will come back to this point later on; for the analytic derivation of this force, see Supporting Information. Also, it is worth highlighting that the perpendicularly polarized pulse in resonance with E1 weakly activates a normal mode in F4TCNQ at 1458~cm$^{-1}$ (${\cal Q}_{d}$ in Figure~\ref{fig:kin_energy}); the same does not happen for a pulse with parallel polarization such as the one stimulating E2 (cf. Figure~\ref{fig:kin_energy}a and b).
A similar behavior, although less pronounced, is exhibited also by the C$\equiv$N modes appearing at frequencies above 2000~cm$^{-1}$ and labeled as ${\cal B}_{1}$ and ${\cal B}_{2}$ in Figure~\ref{fig:kin_energy}a; f or their visualization, see Figure~S4.

Upon the excitation of E2, a different scenario is disclosed, see Figure~\ref{fig:kin_energy}b. 
A significantly larger amount of kinetic energy with respect to the irradiation of E1 is accumulated by ${\cal Q}_{c}$, while the behavior of ${\cal Q}_{b}$ is essentially the same in the two cases.
Moreover, it can be seen that upon resonant excitation of E2, the C$\equiv$N mode at 2218~cm$^{-1}$ (${\cal B}_{2}$) gains kinetic energy throughout the whole simulation window in a more consistent and pronounced way than upon the excitation of E1. 
We explain this behavior considering that the C$\equiv$N bond lies on the plane of the acceptor molecule and, thus, it is aligned with the polarization of the external field in resonance with E2. 
For completeness, it is worth specifying that in 4T-F4TCNQ, the acceptor molecule is slightly bent, with the N atoms pointing toward the donor.
This is a known behavior related to the role of the cyano groups in mediating charge transfer in those organic complexes~\cite{zhu+11cm}. 

Additional insight can be gained by contrasting the power spectrum of the complex against the contributions of its components.
The results displayed in Figures~\ref{fig:kin_energy}c-d represent the amount of energy density accumulated by the system when excited in resonance with E1 and E2, respectively, around the frequency of the corresponding incident pulse (1.2~eV~=~9678.65 cm$^{-1}$ for E1 and 1.5~eV~=~12098.31 cm$^{-1}$ and for E2). 
When the complex is pumped at the energy of E1, the power accumulated due to the excitation of the vibrational degrees of freedom is almost evenly distributed between 4T and F4TCNQ (Figure~\ref{fig:kin_energy}c).
Notice that, in this case, the power spectrum has maxima not only at the laser frequency but also at higher energies.
These additional bands are ascribed to the interaction between the harmonic nuclear motion and the fast oscillating force due to coherence with the field, which is discussed in the Supporting Information.

In the resonant stimulation of E2, on the other hand, the vibronic response of the complex mostly comes from the donor (Figure~\ref{fig:kin_energy}d) with the contribution of the acceptor peaked around the carrier frequency of the incident pulse. 
We recall that in E2, dominated by the HOMO-1$\rightarrow$LUMO transition, the excited charge density is still delocalized over the whole complex but the excitation is polarized along the long molecular axes~\cite{vale-cocc19jpcc,valencia+PCCP+20}. As such, the intramolecular modes couple more efficiently with the electronic dynamics, and the distribution of the kinetic energy is no longer equally spread between the donor and acceptor. 
As shown in Figure~\ref{fig:kin_energy}b, the coupled normal modes that mostly accumulate kinetic energy upon the excitation of E2 are ${\cal Q}_{c}$ and ${\cal B}_{2}$.
On the other hand, the energy accumulated by ${\cal Q}_{a}$ and ${\cal Q}_{b}$ is much weaker compared to the case in which E1 is stimulated, explaining therefore the asymmetric distribution of kinetic energy in favor of the donor (Figure~\ref{fig:kin_energy}d).

\begin{figure*}
\includegraphics[width=\textwidth]{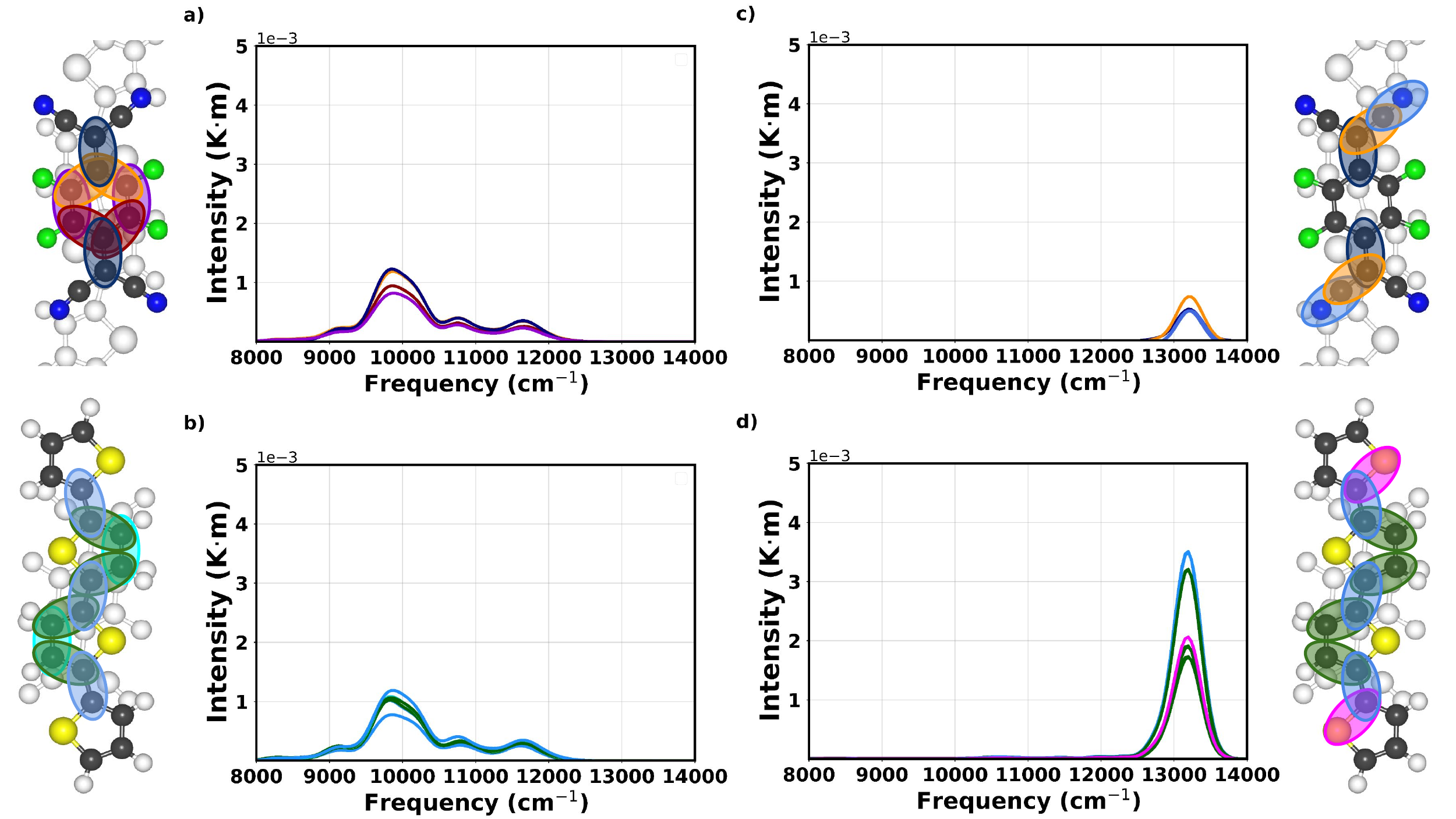}
\caption{\label{fig:ps} Power spectra associated with the dominant vibrational modes of F4TCNQ (panels a and c) 4T (panels b and d), when the complex is excited by a laser pulse in resonance with a)-b) E1 and c)-d) E2. The color code for the solid lines in the spectra shows each bond of the complex by symmetry, and they are shown in the ball-and-stick representation of the complex shown on the side of each panel. 
}
\label{fig:ps_0K}
\end{figure*}

For a deeper comprehension of these results, we now inspect the power spectrum associated with the single bond oscillations in the donor and the acceptor molecule of the CTC excited in resonance with E1 and E2.
In Figure~\ref{fig:ps_0K}, left (right) panel, results obtained upon the excitation of E1 (E2) are shown; on the top and bottom panel, the contributions of vibrations in F4TCNQ and in 4T are reported, respectively.
Upon excitation of E1, the strongest vibronic contribution in the acceptor comes from the C=C mode (Figure~\ref{fig:ps_0K}a).
This is not surprising, considering the participation of this oscillation in the coupled modes ${\cal Q}_{a}$, ${\cal Q}_{b}$, and ${\cal Q}_{c}$ (see Figure~S3).
Additional contributions come from the C-C bonds in the tetrafluorinated phenyl ring.
Looking at the response of 4T to the same excitation, we find the strongest response arising from the C$_{\alpha}$-C$_{\beta}$ and the C$_{\beta}$-C$_{\beta}$ bond oscillations (Figure~\ref{fig:ps_0K}b). 
Notice that among the four most important vibronic contributions shown in Figure~\ref{fig:ps_0K}b, only C-C oscillations appear.
Moving now to the right panels of Figure~\ref{fig:ps_0K}, we appreciate the small contribution of the intramolecular modes of F4TCNQ already discussed with reference to Figure~\ref{fig:kin_energy}d.
The results shown in Figure~\ref{fig:ps_0K}c underline the relevance of the C=C bond oscillation in the acceptor, accompanied, however, by a strong contribution from the C-C bond connecting the cyano group to the phenyl ring.
Notably, the weight of the C$\equiv$N oscillations is larger than the one associated to the phenyl ring. 
The C$_{\alpha}$-C$_{\beta}$ and  C$_{\beta}$-C$_{\beta}$ modes dominate the vibronic response of 4T also upon the resonant excitation of E2 (Figure~\ref{fig:ps_0K}d).

This analysis suggests that the excitation of E1 and E2 by a resonant pulse triggers two different types of vibronic effects. 
For the latter, it is due to nuclear relaxation within the individual components; for E1, they correspond to forced nuclear oscillations driven by the coherence between the ground state and the excited state, facilitated by the relatively low energy of this electronic excitation. 

\subsection{Charge-Transfer Dynamics at Finite Temperatures}

The analysis performed so far on the laser-induced charge-carried dynamics at 0~K equips us with the necessary knowledge to investigate the dynamics of the CTC at finite temperatures. 
Due to the smearing of E1 and E2 in the linear absorption of 4T-F4TCNQ (Figure~\ref{fig:t_density}a), at 100~K, 200~K, and 300~K it is no longer possible to resonantly target the two excitations separately.
For this reason, the results presented below for each temperature value are obtained with a single pulse set at the energy of the lowest-energy peak in the corresponding spectrum in Figure~\ref{fig:t_density}a.

\begin{figure*}
\includegraphics[width=0.5\textwidth]{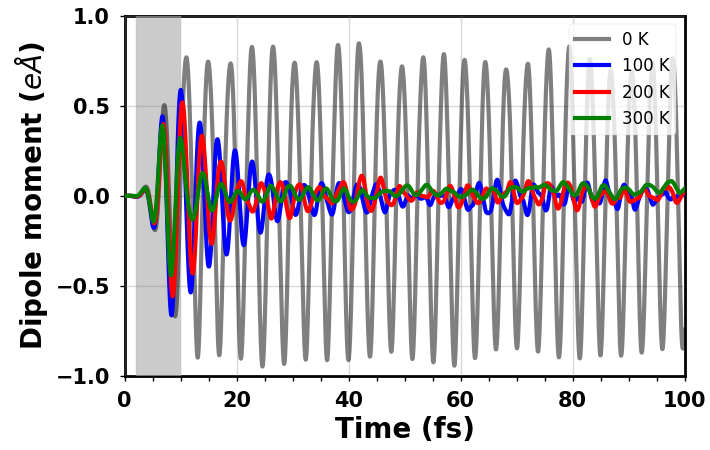}
\caption{\label{fig:ens_dipole} Dipole moment in the $z$-direction induced by the resonant excitation of E1 and averaged over 20 snapshots for each finite temperature. The result at 0~K is shown in gray in the background. The shaded area marks the time window of the pulse.  }
\end{figure*}

Before diving into the discussion of dynamical charge transfer at finite temperatures, we make a short digression to
inspect the $z$-component (across the interface) of the time-dependent induced dipole moment for each ensemble at 100~K, 200~K, and 300~K, and contrast it against the one obtained at 0~K (see Figure~\ref{fig:ens_dipole}).
At zero temperature (gray line), the dipole moment induced by the stimulation of E1 exhibits large oscillations between positive and negative values in a range between -1 and +1~$e$\AA{} over the entire 100-fs window.
These persisting oscillations are a signature of coherence in the absence of any dissipation effect.
Their quantum-mechanical and electronic nature was demonstrated in previous work~\cite{krumland+2020jcp} through a direct comparison between RT-TDDFT simulations without nuclear motion and the results of a two-level model.
Vibrational effects captured here by the applied Ehrenfest formalism manifest themselves in the amplitude modulation of the dipole moment.

Considering now the results for the ensembles averaged at 100~K and 200~K (blue and red curves in Figure~\ref{fig:ens_dipole}, respectively), the maximal amplitude is reached at 12~fs, right after the pulse is switched off. However, differently from the result obtained at 0~K, it decays to less than 0.1~$e$\AA{} within the first 30~fs.
When temperature increases, each nuclear trajectory remains coherent but the evaluation of the macroscopic polarization as an average across all molecules in the ensemble decays due to thermal fluctuations of the excitation energy~\cite{krum+22prb}. The dephasing undergone by the induced dipole moment at finite temperatures is influenced by the specific temperature value.
In particular, in the ensemble at 300~K (green curve), the maximal amplitude is reached at about 7~fs, when the laser is still on, and drops almost immediately after the switch-off of the pulse. 
As a note of caution, it is worth specifying that Ehrenfest dynamics does not obey detailed balance~\cite{para-tull06jctc,liu+22algorithm}. Hence, temperature-induced damping of coherent Ehrenfest dynamics may not necessarily describe the actual physical situation. 

\begin{figure*}
\includegraphics[width=1\textwidth]{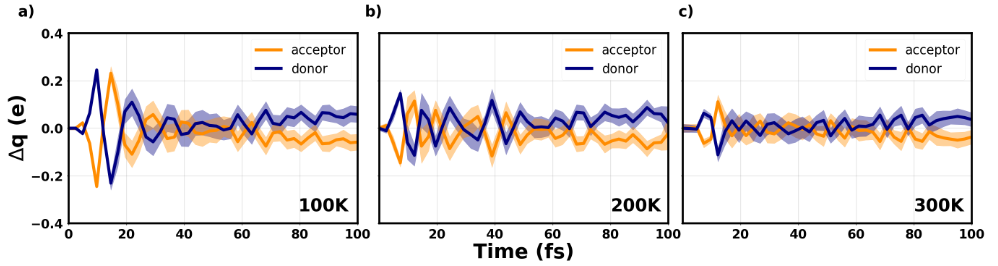}
\caption{\label{fig:ens_bader} Partial charge variation over time with respect to the ground-state value at the considered temperatures of a) 100~K, b) 200~K, and c) 300~K. The results obtained at each time step are averaged over 20 snapshots and the corresponding uncertainty is visualized by the shaded area surrounding the solid lines. 
}
\end{figure*}

In the next step, we investigate how temperature affects the dynamical charge transfer in the laser-excited CTC. 
In analogy with the analysis performed at 0~K, we monitor the temporal variation of the partial charges with respect to the ground-state value obtained at the corresponding temperature (see Table~\ref{tab:table_ct}).
The results shown in Figure~\ref{fig:ens_bader} indicate an oscillatory behavior of $\Delta q$ which assumes positive and negative values both on the donor and on the acceptor throughout almost the entire simulation window. 
In the first 50~fs, the amplitude of the oscillations decreases systematically as the temperature increases: this suggests not only that the charge transfer is less effective at higher temperatures but also that the system consistently loses coherence.
Moreover, it is worth emphasizing once again that at finite temperatures the second excited state, E2, is likely off-resonantly pumped in the stimulation of E1, due to the energetic proximity of these two excitations in the corresponding linear absorption spectra (see Figure~\ref{fig:t_density}a). 
The participation of an off-resonant excitation in the dynamics impacts the coherence.
In the second half of the simulation window, the values of $\Delta q$ are steadily positive on the donor and negative on the acceptor at all considered temperatures (see Figure~\ref{fig:ens_bader}). The time evolution of the resonantly excited thermal ensemble leads to an effective decrease of charge transfer with respect to the corresponding ground-state value.
This result can be understood by recalling that the donor and acceptor molecules in the complex are driven slightly apart from each other (Figure S5) as a concomitant effect of the increasing temperature and of the vibronic response to the external field.

\begin{figure*}
\includegraphics[width=\textwidth]{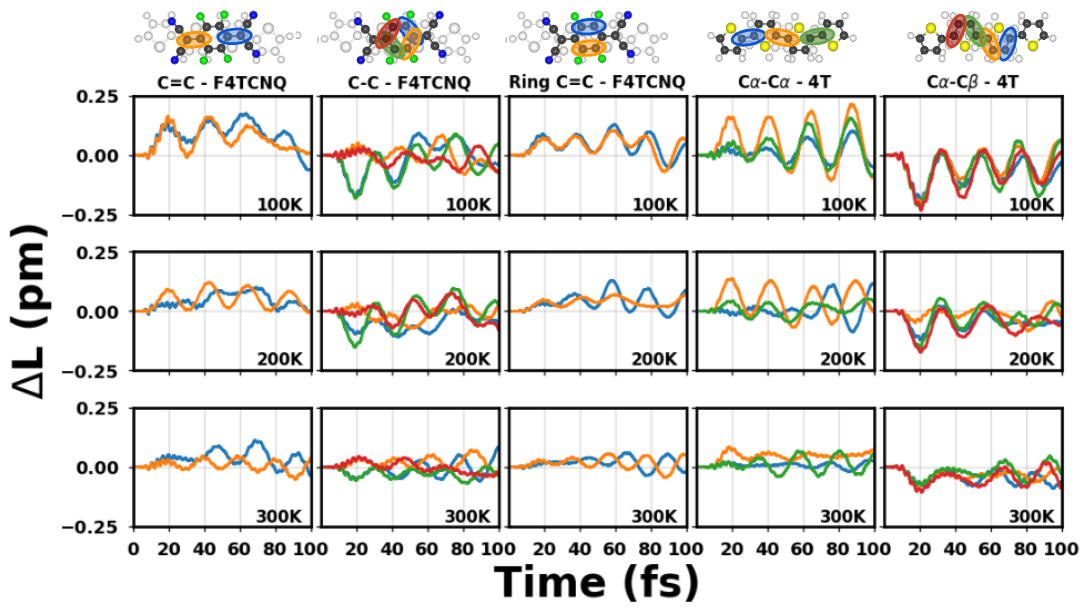}
\caption{\label{fig:ens_bonds} Averaged laser-induced bond length variation ($\Delta L$) with respect to the ground state with and without the pulse for each ensemble thermalized at 100~K, 200~K, and 300~K. 
The bonds involved in the dynamics are highlighted in color in the ball-and-stick sketches of the CTC (top).
}
\end{figure*}

To investigate these effects more in-depth, we monitor the laser-driven time evolution of the bond lengths involved in the vibrational normal modes of the complex (Figure~\ref{fig:ens_bonds}). 
Let us consider first the dynamics of the C=C bond in the acceptor as a function of temperature. 
At 100~K, both bond lengths oscillate in phase until 50~fs, when dephasing starts off. 
At higher temperatures, in-phase C=C oscillations occur only in the first 20~fs; afterward, they begin to dephase. 
An analogous behavior is seen in the dynamics of the C=C bonds in the phenyl ring of F4TCNQ, as well as in the C$_{\alpha}$-C$_{\alpha}$ and the C$_{\beta}$-C$_{\beta}$ oscillations in 4T.
Note that all the corresponding modes mostly participate in the coherent charge-transfer process at 0~K (Figures~\ref{fig:kin_energy} and \ref{fig:ps_0K}). 
The C-C bonds in F4TCNQ are less involved in the laser-induced dynamics: they oscillate in phase in the first few fs but then, at all temperatures, they dephase as the system evolves in time. 
The decreasing amplitude in the bond length oscillations as a function of temperature is associated with the disorder that occurs when temperature increases. In this case, the system can access different regions of the configurational space, and, by doing so, it has different ways to reorganize itself. In other words, more configurations imply higher disorder, while the donor and the acceptor increase their mutual distance as a function of time at finite temperatures (see Figure~S5).
As seen from Figure~\ref{fig:ens_bonds} and Figure~\ref{fig:ens_bader}, this disorder has a detrimental impact on charge transfer. Averaging over many configurations overall weakens the coupling between 4T and F4TCNQ, and hence their ability to dynamically transfer charge across their interface. 
We conclude this discussion by recalling that in the adopted formalism, the zero-point energy is neglected. It is known from the literature that overlooking this effect may influence the observed energy flow as well as the thermalization~\cite{alim+92jcp,shu+18pccp,mukh-barb22jctc}, in particular with regard to the high-frequency modes, which typically are not strongly affected by temperature. 

%%%%%%%%%%%%%%%%%%%%%%%%%%%%%%
\section{Conclusions}
We have presented a first-principles study on the ultrafast dynamics of an organic complex formed by a 4T molecule $p$-doped by F4TCNQ.
Such a system exhibits a fractional degree of charge transfer in the ground state.
We have monitored in time the distribution of the charge carriers following the excitation of the first and second excited states by means of an ultrafast resonant pulse. 
Stimulating the lowest-energy transition between bonding HOMO and the anti-bonding LUMO triggers large-amplitude charge oscillations from the donor to the acceptor, consistent with the system being in a coherent superposition between the ground- and the excited state.
Targeting the second excitation, which is polarized perpendicularly to the interface, \textit{i.e.}, along the long molecular axes, leads to a qualitatively similar to the case of F4TCNQ doping a silicon nanocluster~\cite{jaco+20apx}.
The analysis of the vibrations paired with the photoexcited electronic degrees of freedom sheds light on the fundamental mechanisms ruling the two scenarios. 
We find significant nuclear motion driven by the laser and accompanying harmonic vibrations in spite of the inertia of the heavy nuclei.
Coherently excited vibrational modes coupling the donor and the acceptor molecules are responsible for the dynamics of the lowest-energy excitation; for the second one, intramolecular bond oscillations, especially in the donor, play a more prominent role. 
We have completed our study analyzing temperature effects.
To do so, we have performed Born-Oppenheimer molecular dynamics simulations to obtain thermalized configurations at 100, 200, and 300~K. 
In this scenario, we have found disorder effects becoming predominant and effectively decreasing charge transfer
with respect to the ground state. 

The obtained results provide new insight into the already well-explored area of dynamical charge transfer in organic complexes. Most of the existing work dedicated to the dynamics of organic donor/acceptor complexes was focused on systems forming ion-pairs upon photoexcitation. In the latter systems, the electron-hole pair formed in one subsystem is separated through the mediation of the vibronic motion. In the complex considered here, instead, charge transfer occurs already in the ground state due to the strong p-doping driven by the acceptor. The frontier orbitals are hybridized with bonding and anti-bonding character and the transition between them is an actual charge-transfer excitation with the hole and the electron localized on the donor and the acceptor molecule, respectively. In such a scenario, we analyzed the charge-transfer dynamics and the influence of vibrational and thermal degrees of freedom. Due to the peculiar nature of the investigated system, the obtained results may seem to contradict intuition which is built, however, on systems with substantially different characteristics (ion-pair formation upon photoexcitation) and in which charge transfer is driven predominantly by vibronic coherence~\cite{rozz+13natcom,falk+14sci}.
With our \textit{ab initio} study, we have explored another scenario and discussed the effects of vibronic and thermal motion therein.
We hope that our findings will stimulate corresponding experimental investigations that confirm or challenge them.
In parallel, additional theoretical studies adopting more advanced approaches for the nuclear dynamics, accounting, for instance, for quantum nuclear effects~\cite{you+19es,krum+22prb} may further enhance the understanding of the photophysical properties of charge-transfer complexes subject to intense and coherent irradiation.

%%%%%%%%%%%%%%%%%%%%%%%
%%%%%%%%%%%%%%%%%%%%%%%%%%%%%%%%%%%%%%%%%%%%%%%%%%%%%%%
\begin{acknowledgement}
Stimulating discussions with Mariana Rossi and Antonietta De Sio are kindly acknowledged.
This work was funded by the Deutsche Forschungsgemeinschaft (DFG) - Projektnummer 182087777 - SFB 951 and by the German Federal Ministry of Education and Research (Professorinnenprogramm III) as well as from the State of Lower Saxony (Professorinnen für Niedersachsen, DyNano, and SMART).
Computational resources provided by the North-German Supercomputing Alliance (HLRN), projects bep00076 and bep00104.

\end{acknowledgement}

\begin{suppinfo}
Additional theoretical analysis and \textit{ab initio} results are provided.

\end{suppinfo}


%\bibliography{bib}

\providecommand{\noopsort}[1]{}\providecommand{\singleletter}[1]{#1}%
\providecommand{\latin}[1]{#1}
\makeatletter
\providecommand{\doi}
  {\begingroup\let\do\@makeother\dospecials
  \catcode`\{=1 \catcode`\}=2 \doi@aux}
\providecommand{\doi@aux}[1]{\endgroup\texttt{#1}}
\makeatother
\providecommand*\mcitethebibliography{\thebibliography}
\csname @ifundefined\endcsname{endmcitethebibliography}
  {\let\endmcitethebibliography\endthebibliography}{}

\newpage
%%%%%%%%%%%%%%%
\section*{TOC Graphic}
\begin{figure}[H]
    \centering
    \includegraphics[width=8.25 cm]{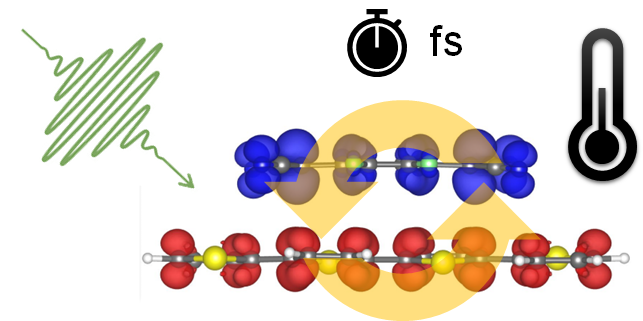}
\end{figure}
%%%%%%%%%%%%%%

\end{document}